\documentclass[nocompress,12pt]{spieman}  
\usepackage{amsmath,amsfonts,amssymb}
\usepackage{graphicx}
\usepackage{setspace}
\usepackage{tocloft}

\title{Speckle Statistics in Adaptive Optics Images at Visible Wavelengths} 
\author{Marco Stangalini\supscr{a,b}, Fernando Pedichini\supscr{a,b}, Enrico Pinna\supscr{d,b},  Julian Christou\supscr{e}, J. M. Hill\supscr{e}, Alfio Puglisi\supscr{d,b}, Vanessa Bailey\supscr{g},  Mauro Centrone\supscr{a}, Dario Del Moro \supscr{c},  Simone Esposito\supscr{d,b}, Fabrizio Fiore\supscr{a,b}, Emanuele Giallongo\supscr{a,b}, Phil Hinz\supscr{f},  Amali Vaz\supscr{f}}

\affiliation{\supscrsm{a}INAF-OAR, Astronomical Observatory of Rome, National Institute for Astrophysics, Italy\\
\supscrsm{b}ADONI Adaptive Optics National Lab of Italy\\
\supscrsm{c}Universit\'a di Roma Tor Vergata, Rome, Italy\\
\supscrsm{d}INAF-Arcetri Astrophysical Observatory of Florence, National Institute for Astrophysics, Italy\\
\supscrsm{e}LBTO, University of Arizona, Tucson AZ 85721, USA\\
\supscrsm{f}CAAO, Steward Observatory, University of Arizona, Tucson AZ 85721, USA\\
\supscrsm{g}KIPAC-Stanford University, Stanford CA, 94305 USA}

\newcommand\apj{ApJ}
\newcommand\apjl{ApJL}
\newcommand\pasp{PASP}
\newcommand\mnras{MNRA}
\newcommand\aap{A\&A}

\cftpagenumbersoff{figure}
\cftpagenumbersoff{table} 
\begin{document} 
\maketitle

\begin{abstract}
{Residual speckles in adaptive optics (AO) images represent a well-known limitation to the achievement of the contrast needed for faint sources detection. Speckles in AO imagery can be the result of either residual atmospheric aberrations, not corrected by the AO, or slowly evolving aberrations induced by the optical system. In this work we take advantage of the high temporal cadence (1 ms) of the data acquired by the SHARK-VIS forerunner experiment at the Large Binocular Telescope (LBT), to characterize the AO residual speckles at visible wavelengths. An accurate knowledge of the speckle pattern and its dynamics is of paramount importance for the application of methods aimed at their mitigation. 
By means of both an automatic identification software and information theory, we study the main statistical properties of AO residuals and their dynamics. We therefore provide a speckle characterization that can be incorporated into numerical simulations \textbf{to increase} their realism, and \textbf{to optimize} the performances of both real-time and post-processing techniques aimed at the reduction of the speckle noise.}  
\end{abstract}


{\noindent \footnotesize{\bf Address all correspondence to}: M. Stangalini, INAF-OAR Via Frascati 33, 00078 Monte Porzio Catone, (RM) Italy; E-mail:  \linkable{marco.stangalini@inaf.it} }

\begin{spacing}{1}   

\section{Introduction}
Speckle noise represents one of the major limitations to the detection of faint companions to nearby stars \cite{1999PASP..111..587R}. Although current high contrast imaging instruments and coronagraphs make all use of sophisticated and high performance AO systems \cite{hugot2012active}, small optical imperfections yield quasi-static speckles and residual stray light, that can still represent a severe limitation to the achievement of the high contrast needed to detect faint companions \cite{marois2000efficient, macintosh2005speckle, cavarroc2006fundamental, 2012SPIE.8447E..0BK}. For these reasons, a deep understanding of the speckle variability and statistics is of fundamental importance \textbf{for the optimization of post-facto techniques aimed at increasing the image contrast \cite{2012A&A...541A.136M, 2013A&A...554A..41M, milli}, like angular differential imaging (ADI), locally optimized combination of images (LOCI), or principal component analysis (PCA) \cite{2006ApJ...641..556M,2007ApJ...660..770L, 2012ApJ...755L..28S, 2012MNRAS.427..948A}}. In addition, it has been demonstrated that speckle intensity statistics represents a powerful tool for speckle discrimination and, therefore, for their post-facto suppression\cite{2010JOSAA..27A..64G}.  
In order to detect a Earth-like planet at close angular distance from a bright star, a flux contrast of the order of $10^{-10}$ is required. Unfortunately, a \textbf{final (after post-processing)} contrast larger than $\sim 10^{-6}$ cannot be currently expected even in AO assisted coronagraphic systems operated under the best seeing conditions at SWIR (Short-wave infrared) wavelengths \cite{2013A&A...554A..41M}. \textbf{Furthermore, an accurate knowledge of the speckle pattern and its dynamics is crucial to increase the realism of simulations aimed at optimizing the instrument performance.}\\

\textbf{The analysis of the dynamical behaviour of AO residuals can be used to estimate the decorrelation time of the atmospheric turbulence and thus its predictability horizon. In Ref.~\citenum{2012SPIE.8447E..0BK} it was recently shown that one of the major limitations to the achievement of the very high contrast needed in exoplanet imaging is represented by the servo-lag error in the ExAO (extreme adaptive optics) systems. Indeed, even a short time lag of $1$ or $2$ ms, can result in a large decrease of the bandwidth of the system, and thus of the overall performances and contrast. In Ref.~\citenum{1992OptL...17..466J} and, more recently, in Ref.~\citenum{2009JOSAA..26..833P} it has been demonstrated that over timescales consistent with the frozen flow approximation, the wavefront aberrations are predictable. In particular, in Ref.~\citenum{2009JOSAA..26..833P} it was shown that $20-40 \%$ of the power spectral density of wavefront fluctuations is due to frozen flow. In order to mitigate the effects of the servo-lag error, several authors have proposed AO control schemes based on different forecasting approaches. See for instance Refs.~\citenum{2004SPIE.5490.1426D, 2010SPIE.7736E..4HS, 2015OptL...40..143J}, to mention a few. Very recently, the validity of this approach has been demonstrated on-sky \cite{tesch2015sky}. The study of the dynamics of AO residuals and their decorrelation time is also useful to investigate the limits of applicability of such techniques.}\\

Several authors have already investigated the statistical properties of speckles in AO corrected images \cite{2004ApJ...612L..85A, 2004EAS....12...89A, 2006ApJ...637..541F, 2011JOSAA..28.1909Y}.
However, as of now, these statistical studies have mostly focused on intensity fluctuations in specific locations of the focal plane, using data sequences with a limited temporal cadence ($>40-50$ ms), longer than the typical atmospheric timescales ($5-10$ ms). \textbf{Recently other authors} \cite{milli} have  investigated the speckle lifetime in the H-band, by exploiting a $1.6$ Hz cadence set of extreme AO (ExAO) images.\\
In order to complement these studies, in this work we exploit new data acquired by the SHARK forerunner experiment \cite{2014SPIE.9147E..8FS, 2014SPIE.9147E..7JF, 2015IJAsB..14..365F, 2016arXiv160905147P} at very high cadence (1 ms), and visible wavelengths. This data allows us to study the behavior of residual AO speckles down to very short timescales, and assess the atmospheric clearance time with very high accuracy. In addition we also exploit these data to study the spatial distribution of long-lived speckles which, as already mentioned, represent a severe limitation to the achievement of very high contrast in ADI images.\\
   \begin{figure}[h]
   \centering
   \includegraphics[width=15cm, clip]{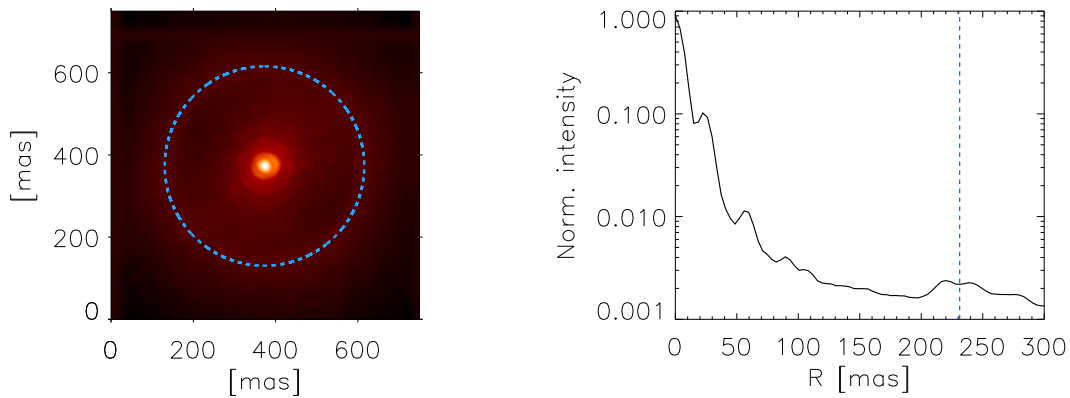}
   \caption{\textbf{\textit{Left panel}: Sum of 1200000 co-registered images ($20$ s equivalent long exposure) of Gliese 777 in logarithmic scale. The  dashed circle indicates the approximate position of the control radius. \textit{Right panel}: Normalized radial profile of the long exposure image. The vertical dashed line indicates the control radius.}} 
    \label{meanimg}
   \end{figure} 

\section{Data set}
\begin{figure*}[!t]
\centering
\includegraphics[width=14cm, clip]{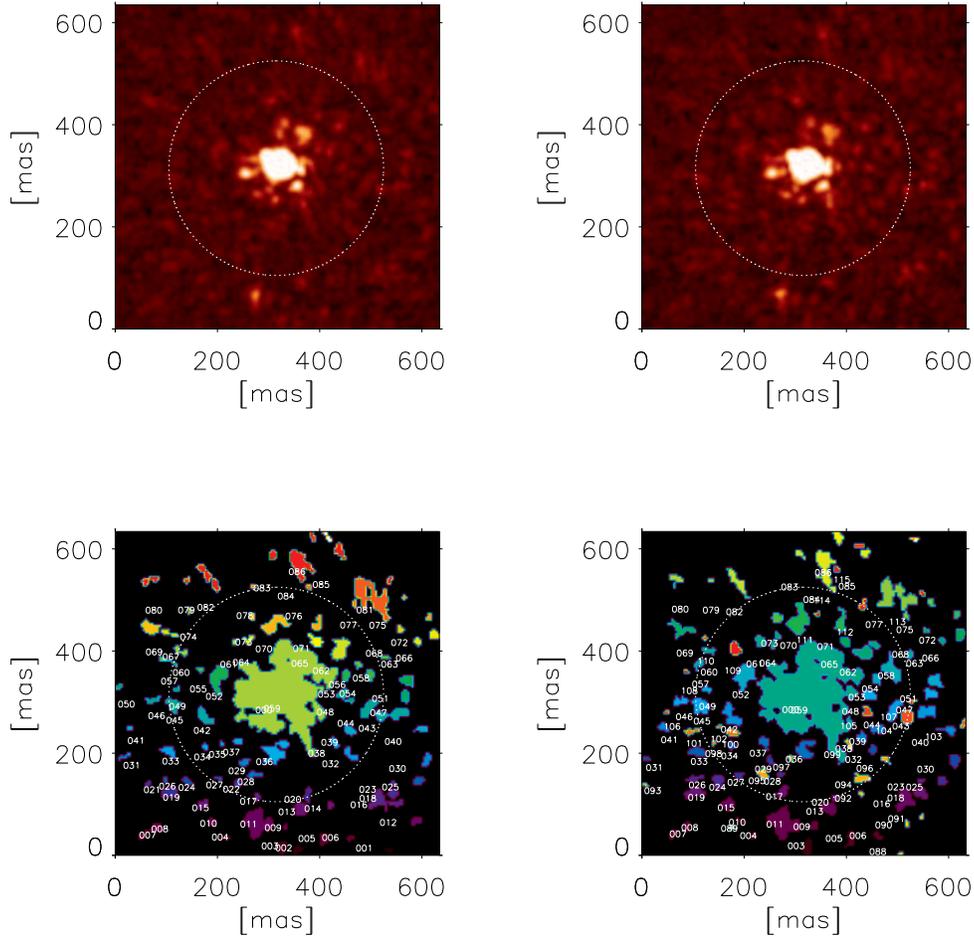}
\caption{\textbf{Two sample logarithmically scaled images}, sequentially acquired every one millisecond (upper panels), and identification of speckles through the SWAMIS code (bottom panels). The speckles are labeled with different colors. Each speckle is assigned a unique identifier throughout the whole data sequence. The white circles indicate the position of the control radius. Only speckles within this radius are considered in the analysis (see text for more details).} 
    \label{imgs}
   \end{figure*}

The data set used in this work consists of a series of 1 ms exposure images of the target Gliese 777 (1 ms cadence), acquired with the SHARK forerunner experiment at LBT on June 4, 2015 (see \textbf{upper panels} of Fig. \ref{imgs}). The pixel scale is set at $3.73$ mas, and the imager is a Zyla CMOS camera manufactured by Andor Inc\footnote{http://www.andor.com/}. The total duration of the data series is $ 20$ min (or equivalently 1200000 images). During the acquisition the LBTI-AO system \cite{esposito2010first} was correcting 500 modes in closed loop. \textbf{The AO frequency was 1 KHz with a loop delay of $3$ ms. In this conditions the closed loop $0$ db bandwidth is $59$ Hz. The seeing $S$ was in the range $0.8 < S < 1.5$ arcsec, and no field de-rotator is employed on the mount to correct for the sky rotation.} \\

The SHARK-VIS Forerunner experiment is a set of short test observations performed at the LBT telescope to verify its AO system performance at visible wavelengths ($600 < \lambda < 900~$ nm) between February and June 2015 \cite{2016arXiv160905147P}. The experimental setup is minimal and composed by only two optical elements before the detector: one divergent lens to get a super sampling (twice the Nyquist limit) of the PSF (point spread function) and a $40$ nm FWHM filter centered at $630$ nm. The AO control and wavefront sensing is left to the LBTI Adaptive Optics subsystem fed through a $50\%$ beam splitter. All the forerunner hardware is placed on a steel optical  breadboard for an easy customization, and attached to the LBT main structure. Data acquisition is performed using an in house developed LabView software interfacing the Andor Zyla sCMOS (Scientific CMOS) camera  with a camera link to a PCI board yielding a maximum throughput of 140 Mpixel/s digitized at $16$ bit.  \textbf{The short exposure time (1 ms)} is used to freeze the evolution of atmospheric speckles and to easily recover the residual jitter in the focal plane, for its post-facto correction. We remark here that residual jitter in longer exposure images may affect the results of the statistical analysis of residual speckles. Our very fast cadence allows us to reduce this effect by employing a post-facto registration of the data sequence.
Our target is acquired at low Zenith angles to avoid PSF elongation due to atmospheric differential refraction, and close to the meridian to maximize the effect of field rotation useful for ADI post processing of image stacks.\\

The data calibration process consists of the dark frame subtraction and image registration through a FFT (Fast Fourier Transform) phase correlation technique. \textbf{In short, the FFT cross correlation between the two images is computed, and the phase is estimated from it. For two perfectly matching images, the cross-phase is a function with a peak at the center of the image domain. When one of the two images is shifted with respect to the other, the position of the peak of the phase function is shifted by an amount of pixels corresponding to the exact shift between the two images themselves. By fitting a gaussian to the phase function it is possible to estimate the shift (i.e. the position of its peak) with sub-pixel accuracy.}\\ 

In the left panel of Fig. \ref{meanimg}, we show an image of the target obtained by \textbf{adding up to 1200000 unsaturated images after their sub-pixel registration. This is equivalent to a 20 min exposure time}. In the same figure (right panel) we also show a radial profile of the same long-exposure PSF.\\
I\textbf{n this work we focus our attention on both regions within and outside the radius at which the AO system is effectively suppressing the atmospheric aberrations.} This corresponds to a distance from the optical axis which is commonly referred to as control radius \cite{davies2012adaptive}, and depends on the number of actuators of the DM (deformable mirror), \textbf{the wavelength, and telescope diameter}. The control radius is represented by the dashed line in Fig. \ref{meanimg}. At larger radial distances the DM is not able to suppress aberration modes, and the PSF is dominated by seeing.\\
\textbf{The speckles within the control radius, are those which are due to either AO residual aberrations, and quasi-static distortions due, for example, to NCPA (non common path aberrations)}. Hereafter we refer to these speckles as \textit{AO residuals}, while we refer to the speckles outside the control radius as \textit{seeing-induced speckles}. However, it is worth mentioning here that \textbf{the} SHARK forerunner is designed to minimize NCPA aberrations by simplifying its optical path and picking the beam off very close to the WFS \cite{2014SPIE.9147E..8FS, 2015IJAsB..14..365F}. 
\textbf{Further, during the observation, the NCPAs aberrations where mitigated by injecting an offset on the secondary deformable mirror of LBT following the procedure described in Ref.~\citenum{esposito2015non}. \\}

\section{Methods and Results}
\subsection{Speckle lifetime statistics}
In order to estimate the lifetime of the two populations of \textbf{speckles within and outside the control radius (i.e. AO residuals and seeing-induced respectively)}, speckles are identified and tracked by using the SWAMIS tracking code \cite{2007ApJ...666..576D}. This code was originally written for the identification and tracking of small scale magnetic elements in the solar photosphere \cite{2008ApJ...674..520L, 2010ApJ...720.1405L, 2013ApJ...774..127L}; a task conceptually similar to that of the analysis of AO residual faint speckles in SHARK forerunner data \cite{2014A&A...561L...6S, 2015A&A...577A..17S}. In short, the code identifies and tracks, through sequential images, small scale features which are above a user specified threshold (\textbf{$4 \sigma$ in our case}), where $\sigma$ is the standard deviation of the signal (intensity) \textbf{computed in a dark region in the upper left corner of the image}, and covering at least an area of \textit{n} pixels (\textbf{where $n=16$ in our case}). \textbf{This implies that only structures with a size of the order of the PSF are identified and tracked, thus ruling out the possibility to include noise features.} In order for a speckle to be uniquely identified, this two thresholds have to be met at the same time.\\ 

\textbf{With the aim of reducing the computational time, instead of using the entire set of images (1200000), we limit ourselves to the analysis of different temporal windows of 5 s (5000 images) during the observation run}. 
Indeed, even if looking at this short interval, the number of speckles identified and \textbf{matching the above criteria amounts to about $90000$}. This is a large number ensuring a good estimation of the underlying statistics. \\
\textbf{Here we focus our attention to the first temporal window at the beginning of the data sequence, while at the end of this section we will extend the analysis to the other temporal windows.\\}
After the identification of the speckles matching the searching criteria, the code assigns a label to each of them. This label is unique throughout the data series and permits the identification of the speckles at different time steps. In the lower panels of Fig. \ref{imgs}, we show an example of masks obtained from the identification of the speckles in the images shown in the upper panels of the same figure. \textbf{In the same panels, we} also show the labels of each speckle, which represents its unique identifier during its entire life.\\ 

In order to distinguish possible differences between the two populations of speckles, we separate them into two classes: those lying within the control radius, and those outside (see blue dashed line in Fig. \ref{meanimg}), and measure, for each of them, the lifetime and the maximum intensity throughout their entire life.\\
\textbf{It is worth noting that the selection criteria described above do not allow the code to single out speckles in proximity of the center of the PSF (see lower panels of Fig. \ref{imgs}). Despite this, a large amount of speckles are identified within the control radius anyway, thus ensuring the statistical significance of the results.}

\begin{figure*}[t!]
\centering
\includegraphics[width=16cm, clip]{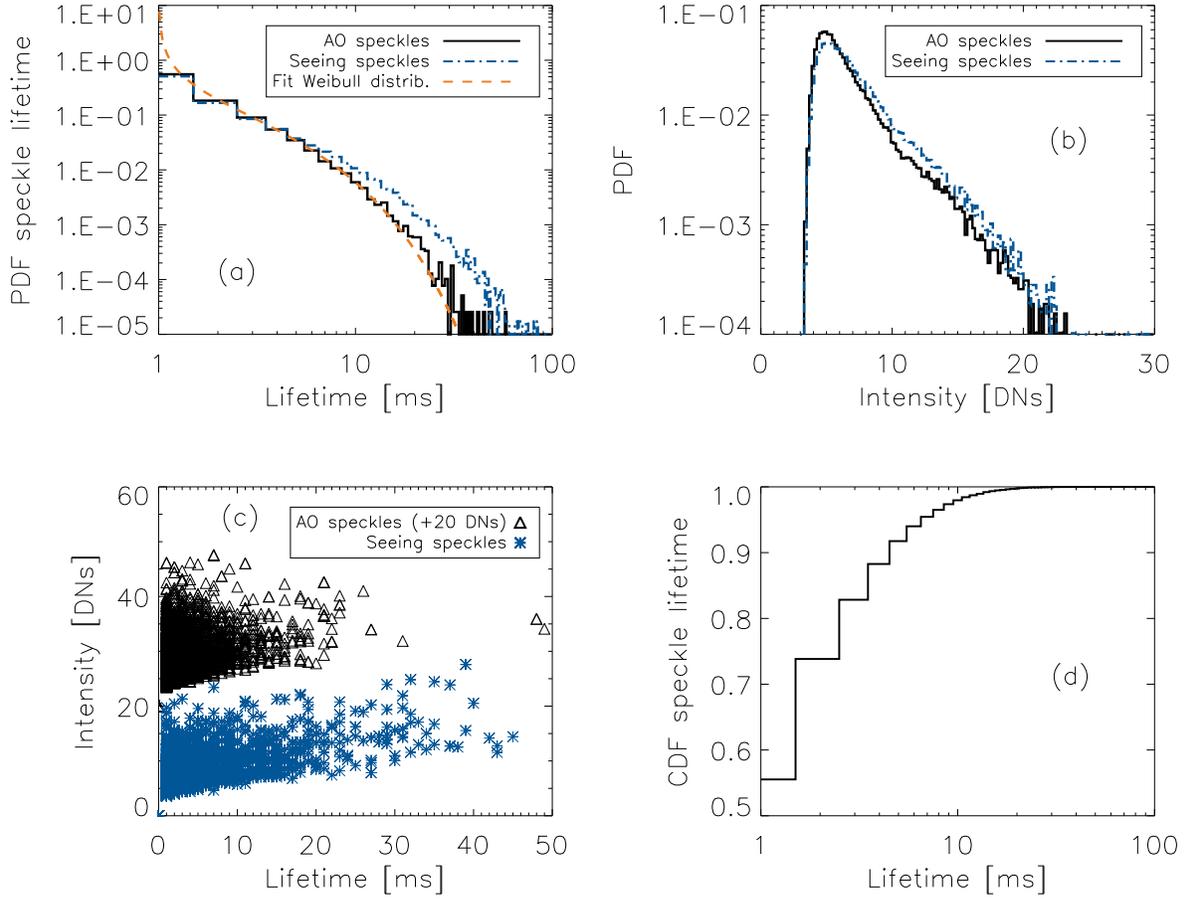}

\caption{\textbf{\textit{panel a}: PDF of the lifetimes of the AO and seeing speckles. The orange line represents the least squares fit to the data of a Weibull distribution. \textit{panel b}: PDF of the intensity of both populations of speckles. \textit{panel c}: Lifetime-intensity relation for the two samples. Please note that the intensity of the AO residual speckles was shifted upwards by an amount equivalent to $10$ DNs for graphical reasons. \textit{panel d}: CDF} (cumulative distribution function) \textbf{of the AO speckle lifetime.}} 
    \label{PDFlife}
   \end{figure*}

In panel (a) of Fig. \ref{PDFlife} we show the probability density function (PDF) of lifetimes for the two samples of speckles. As one can note from the same figure, the PDF can be modeled by a 2-parameter Weibull distribution \cite{weibull1951wide} of the form:
\begin{equation}
f(t)=\frac{\beta}{\eta} \left( \frac{t}{\eta} \right)^{(\beta-1)}exp{\left(-	\frac{t}{\eta} \right)^{\beta}},
\end{equation}
where \textbf{$t$ is the time}, $\beta$ a shape parameter, and $\eta$ a scaling parameter, also known as characteristic time. The least \textbf{squares}\textbf{ fit to the data yields $\beta=0.25$ and $\eta=1$ ms}. It is worth recalling that, by definition, the $\eta$ time scale of the Weibull distribution represents the time at which $\sim 63.2 \%$ of the sampled population die.\\
Apart \textbf{from the physical meaning }(Weibull distributions are usually found in turbulence as the signature of random multiplicative processes \cite{frisch1997extreme}), this represents a simple form that can be easily incorporated into numerical simulations of high contrast coronagraphic imagers, to improve the accuracy of their \textbf{estimated  contrast.}\\

It is worth noting that the PDF of the lifetimes of seeing-induced speckles, as compared to that of the AO residuals, shows a larger density of elements in the range $20-70$ ms, if compared to the PDF of the AO residuals. \textbf{This is in agreement with previous results in literature showing that the AO system only modifies the intensity of speckles, leaving their lifetime unchanged\cite{macintosh2005speckle}. Indeed, this effect can be due to the intensity threshold used for the identification of the speckles.   In panel (b) of the same figure we show that the PDF of the intensity of the seeing-induced speckles presents an intrinsically larger amount of elements, with respect to the intensity distribution of AO residuals, in the bright end.} In the presence of an almost linear dependence of the lifetime on the intensity  (see panel c of the same figure), this translates into an increase of the probability of detection of long-lived speckles outside the control radius of the PSF. Brighter speckles remain above the \textbf{selection} threshold longer than the fainter ones, resulting in an increase of the fraction of the \textbf{identified} longest-lived speckles themselves. However, \textbf{except for} this increase, the two PDFs do not show significant differences. \\

\begin{figure*}[t!]
\centering
\includegraphics[width=9cm, clip]{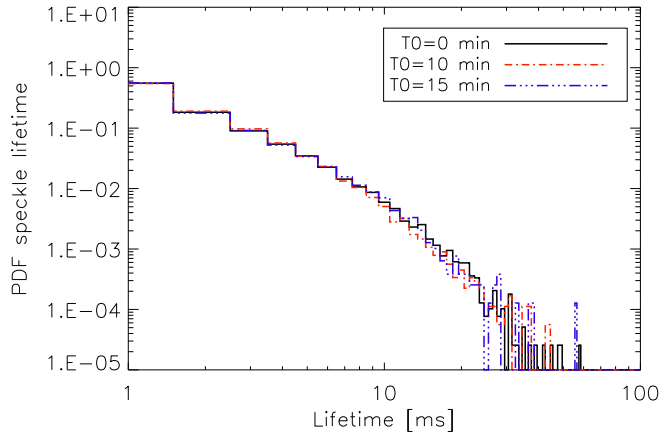}
\caption{\textbf{PDF of the AO speckles lifetime computed in different temporal windows (6 s each) along the entire data sequence.}} 
\label{PDFlife_comparison}
\end{figure*}

Another important aspect by far \textbf{is} the estimation of the atmospheric refreshing time, that is the time over which the speckle pattern is completely renewed.\\
To this regard, in panel (b) of Fig. \ref{PDFlife} we show the CDF (cumulative distribution function) of the AO speckle lifetime. The CDF represents the probability of finding a speckle with a lifetime $\tau_{spk}$ shorter than $\tau$. The CDF shows that while $~50 \%$ of the speckles identified have a lifetime of the order of $1$ ms, $99 \%$ of them are found within $20$ ms. This means that after this time, the speckle pattern is almost completely renewed, and this value can be considered as the refreshing time scale of the atmospheric turbulence.\\
\textbf{The above results were obtained by analyzing a temporal window of $5$ s at the beginning of the data sequence ($T_{0}=0$ min). In order to be sure that the statistics does not depend on the particular temporal window selected, as already mentioned we repeated the analysis over different temporal windows (i.e. $T_{0}=0$, $T_{0}=10$ min, and $T_{0}=15$ min). The results of this analysis are shown if Fig. \ref{PDFlife_comparison} where we plot the PDFs of the speckle lifetime for each temporal window selected. This plot shows that the PDF of lifetimes does not change significantly during the observation.}

\subsection{\textbf{Mutual information and memory of the process}}
In order to independently check the above results, we make use of a completely different approach based upon information theory \cite{shannon2001mathematical}. More in particular, we make use of the mutual information (MI) \cite{paninski2003estimation, kraskov2004estimating}, which is a measure of the non-linear mutual dependence of two variables $X$ and $Y$, and defined as:
\begin{equation}
MI(X,Y)=\sum_{y \epsilon Y} \sum_{x \epsilon X} p(x,y) ~ log \left( \frac{p(x,y)}{p(x)p(y)} \right),
\end{equation}
where $p(x,y)$ represents the joint probability function, and $p(x)$ and $p(y)$ the probability density functions of $X$ and $Y$, respectively. It is a well known result that, while the correlation is a measure of the linear dependence between two variables, mutual information is a more general quantity that can be applied also to non-linear processes \cite{1990JSP....60..823L}. MI was already employed in the adaptive optics field, in the optimization of the wavefront reconstruction \cite{stangalini2010zernike}.\\
\begin{figure}[t!]
\centering
\includegraphics[width=9cm, clip]{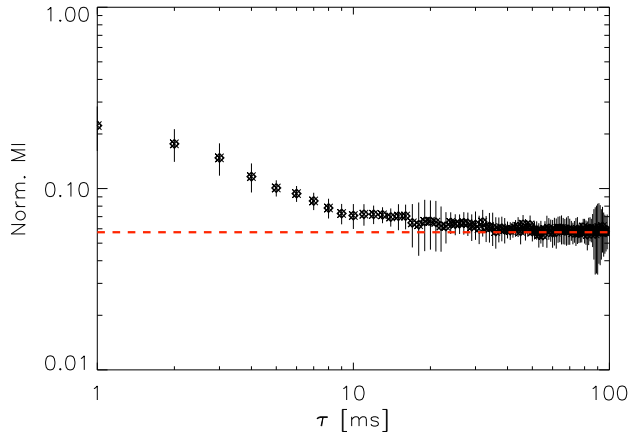}
\caption{Mutual information as a function of the time delay. The error bars represent the standard error of the mean of the $50$ sequences of $100$ ms over which the MI is computed. The MI is estimated for each image with respect to the beginning of the sequence. The horizontal dashed line represents the asymptotic value of the MI estimated as the average value between $80$ and $100$ ms.} 
\label{MI}
\end{figure} 
\textbf{Since the statistics of speckles does not change during the observation, here we focus on the first temporal window analysed at the beginning of the previous section.
} We divided the first $5$ s of the observations \textbf{into} short subsequences of $100$ ms each. \textbf{This is possible since in $100$ ms the speckle pattern is completely refreshed.}  For each subsequence, we estimated the MI between the first and any other following image. This is done by only considering the region of the FoV within the control radius which is marked by \textbf{the} dashed blue line of Fig. \ref{meanimg}. In Fig. \ref{MI} we plot the average MI as a function of the delay $\tau$. We stress here that this analysis is only made possible by the high cadence of the data. As expected, MI undergoes a rapid decrease as the time delay goes by. This indicates a rapid decrease of correlation of the process with time. After $\sim 20$ ms, MI reaches an  asymptotic value, indicating the presence of a persistent pattern of residues. Most of the speckles have a very short lifetime (shorter than $20$ ms) and, consequently, the MI drops by almost $80 \%$ in $1$ ms ($MI=1$ for $\tau=0$ by definition). The rapid reduction of MI reflects the short memory of the process associated with the evolution of the atmospheric turbulence, which determines a fast evolution of the speckle pattern. Indeed, after $20$ ms the speckle pattern is completely renewed, and the mutual dependence between the current and the first PSFs is reduced at minimum. However, it is worth noting that the asymptotic value of the MI is not zero, as expected for statistically independent images. As already anticipated, this implies that there \textbf{exists} a quasi-static component of the pattern itself. However, this component accounts for only $\sim 6 \%$ of the total information contained in the speckle images. We note that the estimated value of the decorrelation time obtained through MI, is in good agreement with that estimated by the PDF of the lifetimes seen in the previous section.\\
\textbf{It is interesting to note that the error bars in the plot, representing the standard errors of the mean of the $50$ sequences, oscillate in size reflecting the intrinsic spread of the MI estimated from the different sequences.}\\

\subsection{Spatial distribution of quasi-static speckles}   
\begin{figure}[]
   \centering
   \includegraphics[width=9cm, clip]{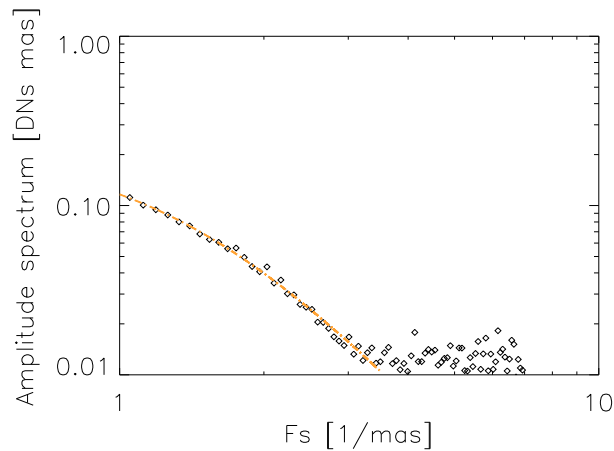}
   \caption{FFT amplitude spectrum of the quasi-static speckles \textbf{as a function of the spatial frequency $F_{s}$}. The dashed line represents the fit of the TPL \textbf{(Terrestrial Planet Finder)} model to the data points.} 
    \label{SpatialFFT}
   \end{figure}

In order to further investigate the nature of AO residual speckles, we also analyse their spatial distribution by estimating the FFT spatial spectra of intensity fluctuations. More in particular, we focus our attention on the spatial distribution of the longest-lived components of the intensity fluctuations with the aim of investigating the character of quasi-static speckles. If images are averaged over a sufficiently longer time (say $7-10$ times the clearance time), we expect to significantly reduce the contribution from seeing-induced and rapidly disappearing speckles, with only the ones commonly referred to as quasi-static speckles left in the images. In order to study the spatial contribution of this latter component, we average \textbf{our entire data sequence (20 min duration)} in windows of $200$ ms to get rid of the rapidly evolving speckle component. For each of this "long exposure" images, we estimate the spatial FFT spectrum, after masking out \textbf{pixels} outside the control radius, over which the DM has no effect. The masking is performed with a gently decreasing function without sharp edges to avoid side effects in the FFTs ($1-H(x,y)$, where $H(x,y)$ represents the Hanning window centered at the PSF peak). We then consider the average amplitude spectrum of the spatial fluctuations as our best estimate of the quasi-static speckles spatial distribution in the Fourier space.
\textbf{The spatial spectrum of the fluctuations due to quasi-static speckles can be modeled as follows \cite{Church1990TheMeasurements}:}
\textbf{\begin{equation}
P(F_{s})=\frac{A_{0}}{[1+(F_{s}/A_{1})]^\frac{A_{2}+1}{2}},
\end{equation}
where $F_{s}$ represents the spatial frequency.\\}
In Fig. \ref{SpatialFFT} we show the mean amplitude of the spatial spectrum of residual speckles. A least \textbf{squares} fit of the above model to the data in the range \textbf{$1-3$ mas$^{-1}$} yields the following estimates for the three parameters: $A_{0}=0.2 \pm 0.04$ $DNs \cdot mas$, $A{1}=1.40 \pm 0.05$ $mas^{-1}$, and $A_{2}=2.0 \pm 0.2$. 

\section{Concluding remarks}
Our results show that the turbulence clearance time at visible wavelength is of the order of $~20$ ms. Indeed, $99.9 \%$ of the statistical sample of speckles detected in our data have a lifetime shorter than this value. \textbf{The CDF} of the speckles lifetime shows that those with a lifetime longer than $\sim 70$ ms are very few; only $0.15$ speckle/s. In addition, the PDF of the lifetime of the AO residual speckles is found to be well represented by a Weibull distribution. This is not surprising since this distribution is generally used to model lifetimes of different physical processes in a wide range of contexts \cite{weibull1951wide}. However, it is worth stressing here that the accurate modeling of the PDF of lifetimes is only made possible by \textbf{the high frame rate} of the SHARK-VIS forerunner experiment that, delivering images with a cadence of 1 ms, freezes the atmospheric turbulence evolution. The clearance time of the atmospheric turbulence is estimated in this work through two different approaches (automatic speckle identification and information theory) from closed-loop AO images \textbf{and} can be regarded as a decorrelation time scale of the spatial pattern of speckles. After this time, the speckle pattern is completely regenerated.\\

\textbf{Our results demonstrate that at least in these particular observing conditions, the decorrelation time of the atmospheric aberrations is much longer than the AO correction frequency (1 kHz).} It is worth noting here that $90 \%$ of the speckles have a lifetime shorter than 5 ms, thus the overall predictability horizon of the wavefront aberrations must lie in this range. In other words, \textbf{at visible wavelengths}, most of the memory of the system is already lost after 5 ms, thus this sets an intrinsic limit to the predictability of the process which is much shorter than the decorrelation time. \textbf{This is important for the implementation of AO predictive control schemes in the visible.}\\

In addition to this, we also estimated the spatial distribution of the quasi-static speckles, after averaging out the seeing-induced ones. The power spectrum of the quasi-static speckles is well represented by the model proposed in Ref. ~\citenum{Church1990TheMeasurements}. However, in this work we are able to provide an accurate estimate of their parameters thanks to the high cadence of our data.\\ \textbf{This piece of information is important to increase} the realism of end-to-end AO simulations for the assessment of the performances of high contrast imagers in real on-sky conditions.\\
It is worth noting that our characterization of the speckle statistics is based on a particular set of observing conditions. However, we also note that this is the first time that seeing induced residual speckles are characterized down to time scales as small as 1 ms. This is due \textbf{to the high acquisition cadence} of the SHARK forerunner that allowed us to perform a detailed characterization of the dynamics of the speckles themselves.\\

\acknowledgments
The LBT is an international collaboration among institutions in the United States, Italy and Germany. LBT Corporation partners are: Istituto Nazionale di Astrofisica, Italy; The University of Arizona on behalf of the Arizona Board of Regents; LBT Beteiligungsgesellschaft, Germany, representing the Max-Planck Society, The Leibniz Institute for Astrophysics Potsdam, and Heidelberg University; The Ohio State University, and The Research Corporation, on behalf of The University of Notre Dame, University of Minnesota and University of Virginia. This work was partially funded by ADONI, the ADaptive Optics National laboratory of Italy, and by the European Commission's H2020 project GREST, grant agreement n. 653982.

\bibliographystyle{spiejour}   


%
%

\vspace{1ex}
\noindent Biographies and photographs of the other authors are not available.


\end{spacing}
\end{document}